\documentclass[proceedings]{stacs}
\stacsheading{2009}{195--206}{Freiburg}
\firstpageno{195}

\usepackage{amsmath,amsfonts, amssymb, xcolor}
\newcommand{\Ac}{\mathcal{A}} 
\newcommand{\Ad}{\mathcal{B}}
\newcommand{\Az}{\mathcal{A}_0} \newcommand{\locA}{\delta_{\Ac}} 
\newcommand{\globA}{G_{\Ac}} \newcommand{\globAd}{G_{\Ad}} \newcommand{\locAz}{\delta_{\Az}}

\newcommand{\Alphh}{\mathcal{Q}} \newcommand{\Alpha}[1]{\Alphh_{#1}}
\newcommand{\Alphn}{\Alpha{n}}  \newcommand{\AlphAd}{\Alpha{\Ad}}
\newcommand{\Alphz}{\Alpha{n_0}}
\newcommand{\ZZ}{\mathbb{Z}} \newcommand{\NN}{\mathbb{N}}
 
\newcommand{\CA}{\textbf{CA}} 
 \newcommand{\CAe}[2]{\textbf{CA}_{#1,#2}}
\newcommand{\CAnk}{\CAe{n}{k}} 
\newcommand{\MS}{\textbf{MS}} \newcommand{\MSnk}{\textbf{MS}_{n,k}}
\newcommand{\Pro}{\mathcal{P}}
\newcommand{\Path}{\rho}
\newcommand{\TOTnk}{\textbf{Tot}_{n,k}}
\newcommand{\SETnk}{\textbf{Set}_{n,k}}
\newcommand{\SET}{\textbf{Set}}
\newcommand{\TOT}{\textbf{Tot}}
\newcommand{\KSET}{\textbf{KSet}}
\newcommand{\KSETnk}{\textbf{KSet}_{n,k}}
\newcommand{\MSK}{\textbf{KMS}}
\newcommand{\oms}[2]{\textbf{O}_{#1}\textbf{MS}_{#2}}
\newcommand{\oset}[2]{\textbf{O}_{#1}\textbf{Set}_{#2}}
\newcommand{\otot}[2]{\textbf{O}_{#1}\textbf{Tot}_{#2}}
\newcommand{\SSnk}{\textbf{SS}_{n,k}}
\newcommand{\Knk}{\textbf{K}_{n,k}}
\newcommand{\Kap}{\textbf{K}}
\newcommand\bloc[1]{b_{#1}}
\newcommand\debloc[1]{b^{-1}_{#1}}

\newcommand\sac{\sqsubseteq}
\newcommand{\simu}{\preccurlyeq}
\newcommand\grp[2]{{#1}^{<#2>}}
\newcommand\fst{\pi_1}

\newcommand{\Nn}{\mathbb{N}}

\newcommand{\focus}[1]{\paragraph{\textbf{#1}}}

\newcommand{\spr}{\#}

\newcommand{\MSC}{\MSK} \newcommand{\mapv}{\varphi}

\newcommand\codms{\Psi}

\newcommand{\Class}{\mathcal{F}}

\newcommand{\Size}[1]{|#1|} 

\newcommand{\Li}[1]{\textbf{L}_{#1}}

\newcommand{\Mapset}{\mathcal{I}}
\newcommand{\ClassE}{\textbf{E}}
\newcommand{\ms}[1]{\{\hspace{-3pt}\{#1\}\hspace{-3pt}\}}
\newcommand{\dlim}[3]{d_{#1,#2}(#3)}
\newcommand{\Simz}{\mathcal{S}_{\Az}}
\newcommand{\Simzx}[1]{\mathcal{S}_{\Az,#1}}
\newcommand{\Univ}{\mathcal{U}}
\newcommand{\Classnk}{\mathcal{F}_{n,k}}
\newcommand{\N}[1]{\mathbb{N}_{#1}}

\newcommand{\0}{0_0}
\newcommand{\1}{1_0}
\newcommand{\K}{\Kap}
\newcommand{\ie}[2]{ [\hspace{-1ex}[\hspace{0.5ex}#1;#2 \hspace{0.5ex} ]\hspace{-1ex}]}
\newcommand{\ies}[2]{ [\hspace{-0.2ex}[\hspace{0.5ex}#1;#2 \hspace{0.5ex} ]\hspace{-0.2ex}]}

\newenvironment{proo}{\proof}{\qed}

\begin{document}

\title[On Local Symmetries and Universality in Cellular
Automata]{On Local Symmetries and Universality in Cellular Automata}

\author[lablama]{L. Boyer}{Laurent Boyer} %
\address[lablama]{LAMA (CNRS, Universit\'e de Savoie),\newline Campus
  Scientifique, 73376 Le Bourget-du-lac cedex FRANCE} 
\urladdr{http://www.lama.univ-savoie.fr} 
\email[L. Boyer]{laurent.boyer@univ-savoie.fr} 
\author[lablama]{G. Theyssier}{Guillaume Theyssier}
\email[G. Theyssier]{guillaume.theyssier@univ-savoie.fr} 




\begin{abstract}
  \noindent
  Cellular automata (CA) are dynamical systems defined by a finite
  local rule but they are studied for their global dynamics. They can
  exhibit a wide range of complex behaviours and a celebrated result
  is the existence of (intrinsically) universal CA, that is CA able to
  fully simulate any other CA. In this paper, we show that the
  asymptotic density of universal cellular automata is 1 in several
  families of CA defined by local symmetries. We extend results
  previously established for captive cellular automata in two
  significant ways. First, our results apply to well-known families of
  CA (e.g. the family of outer-totalistic CA containing the Game of
  Life) and, second, we obtain such density results with both
  increasing number of states and increasing neighbourhood. Moreover,
  thanks to universality-preserving encodings, we show that the
  universality problem remains undecidable in some of those families.
\end{abstract}

\maketitle

\section*{Introduction and definitions}\label{sec:intro}

The model of cellular automata (CA) is often chosen as a theoretical
framework to study questions raised by the field of complex
systems. Indeed, despite their formal simplicity, they exhibit a wide
range of complexity attributes, from deterministic chaos behaviours
(\textit{e.g.}  \cite{Kurka97}) to undecidability in their very first
dynamical properties (\textit{e.g.}  \cite{Kari94}). One of their
most important feature is the existence of universal CA. Universality
in CA is sometimes defined by an adaptation from the model of Turing
machines and sequential calculus. But a stronger notion, intrinsic to
the model of CA, has emerged in the literature \cite{surveyOllinger}:
a CA is \emph{intrinsically universal} if it is able to fully
simulate the behaviour of any other CA (even on infinite
configurations).

Besides, when it comes to modelling \cite{chopard} or experimental
studies \cite{Wolfram83, Wolfram84}, most works focus on some
particular syntactical families (elementary CA, totalistic CA, etc),
either to reduce the size of the rule space to explore, or to match
hypothesis of the studied phenomenon at microscopic level
(\textit{e.g.}  isotropy).

In a word, CA are known for their general ability to produce complex
global behaviours, but local rule considered in practice are often
very constrained. This paper studies the link between syntactical
restriction on CA local rules and typical global behaviours
obtained. It establishes a probabilistic result: for various symmetry
criterions over local rules, randomly choosing a local rule within the
symmetric ones yields almost surely universal CA. Meanwhile, the
universality problem is shown to remain undecidable even restricted to
symmetric rules (for some of the symmetry criterions).

A family of CA defined by a simple syntactical constraint (namely
\emph{captive} CA) and containing almost only universal CA has already
been proposed by one of the authors \cite{Theyssier05}, but the present paper
goes further. First, it generalises the probabilistic framework: the
neighbourhood of CA is no longer fixed as it was needed in
\cite{Theyssier05}. Second, it considers well-known families of CA
(\textit{e.g.} totalistic or outer-totalistic CA) and generalisations
of them, namely \emph{multiset} CA, which are meaningful for
modelling (they are 'isotropic' CA).

After having recalled standard definitions about CA (end of this
section), section~\ref{sec:locsym} presents the families
considered in this paper. Then, section~\ref{sec:simuniv} defines
intrinsic universality and the simulation relation involved in that
notion. Section~\ref{sec:densprop} gives the probabilistic setting of
the paper and establishes the main probabilistic results. Finally,
section~\ref{sec:univexist} is dedicated to existence proofs of
universal CA in the families considered. Combined with probabilistic 
results, it proves that almost all CA are universal in those families.

\focus{Definitions and notations}

In this paper, we adopt the setting of one-dimensional cellular
automata.  
Formally, a CA is a 3-uple ${\Ac=(n, k, \locA)}$ where $n$ and $k$ 
are positive integers, respectively the size of the state set 
${\Alphn = \{1,\ldots,n\}}$ and of the neighbourhood 
${\ie{-\lfloor\frac{k-1}{2}\rfloor}{\lfloor\frac{k}{2}\rfloor}}$, 
$\locA:\Alphn^{k}\rightarrow\Alphn$ is the \emph{local transition function}.

A coloring of the lattice $\ZZ$ with states from $\Alphn$
(\textit{i.e.}  an element of $\Alphn^\ZZ$) is called a
\emph{configuration}.  To $\Ac$ we associate a global function
$\globA$ acting on configurations by synchronous and uniform
application of the local transition function.  Formally, $\globA:
\Alphn^\ZZ\rightarrow\Alphn^\ZZ$ is defined by: ${\globA(x)_z =
  \locA(x_{z-\lfloor\frac{k-1}{2}\rfloor},\ldots,x_{z+\lfloor\frac{k}{2}\rfloor})}$ for all ${x\in\Alphn^\ZZ}$ and ${z\in\ZZ}$.

The local function $\locA$ naturally extends to $\Alphn^\ast$, the
set of finite words over alphabet $\Alphn$ (with ${\locA(u)}$ being
the empty word if ${|u|<k}$). For ${p\in \NN}$, this function maps an 
element of $\Alphn^{p+k}$ to an element of $\Alphn^{p+1}$.

The size of $\Ac=(n,k,\locA)$ is the pair $(n,k)$. 
The set of all CA is denoted by $\CA$, and the set of all CA of size 
$(n,k)$ by $\CAnk$. Moreover for any set $\Class \subseteq \CA$, 
$\Classnk$ is defined by $\Classnk=\Class \cap \CAnk$. 
Formally a CA is a 3-uple but, to simplify notation, we
sometimes consider that $\Classnk$ is a set of local functions of type
$\Alphn^k \rightarrow \Alphn$.

This paper will intensively use (finite) multisets. A multiset $M$
of elements from a set $E$ is denoted by
${M=\ms{(e_1,n_1),\ldots,(e_p,n_p)}}$ where a pair $(e_i,n_i)\in
E\times\Nn$ denotes an element and its multiplicity. The cardinality
of $M$ is $\Size{M}={\sum_i n_i}$. The cardinality notation is the 
same for sets.

\section{Families of CA with Local Symmetries}
\label{sec:locsym}

In this section, we define various families of CA characterised by
some local symmetry. 'Symmetry' must be taken in a broad sense since
it may concern various aspects of the local function. We first
consider families where the local function does not depend on the
exact configuration of the neighbourhood (a $k$-uple of states) but
only on a limited amount of information extracted from this
configuration.

\focus{MultiSet CA} Multiset cellular automata are cellular automata
with a local rule invariant by permutation of neighbours.
Equivalently, they are CA whose local function depends only on the
multiset of states present in the neighbourhood. Formally, $\Ac \in
\CAnk$ is \emph{multiset}, denoted by $\Ac \in \MSnk$,
if for any permutation $\pi$ of ${\{1\ldots k\}}$, the local function
$\locA$ satisfies
\[\forall a_1,\ldots,a_k\in\Alphn : \locA(a_1,\ldots,a_k)=\locA(a_{\pi(1)},\ldots,a_{\pi(k)}).\]

\focus{Set CA} Set CA are a special case of multiset CA: they
are CA whose local function depends only on the set of states present
in the neighbourhood. Formally, $\Ac \in \CAnk$ with arity $k$ is a
\emph{set CA}, denoted by $\Ac \in \SETnk$, if
\[\forall u,v\in\Alphn^k : \{u_1,\ldots,u_k\}=\{v_1,\ldots,v_k\}\Rightarrow\locA(u)=\locA(v).\]
Note that for fixed $n$, there is a constant $N$ such that, for all $k$,
${\Size\SETnk\leq N}$. Thus there is no hope that the asymptotic density
of a non-trivial property for fixed $n$ be $1$ for family $\SET$.

\focus{Totalistic CA} Totalistic CA are also a special case of
Multiset CA: they are CA whose local functions depends only on the
sum of the neighbouring states. Formally, $\Ac \in \CAnk$
$k$ is \emph{totalistic}, denoted by $\Ac \in \TOTnk$, if
\[\forall u,v\in\Alphn^k : \sum_{i=1}^{k}u_i=\sum_{i=1}^{k}v_i\Rightarrow\locA(u)=\locA(v).\]

\focus{Partial Symmetries} We can consider weaker forms of each family
above, by excluding some neighbours from the 'symmetry' constraint and
treating them as a full dependency in the local function. For
instance, we define the set of \emph{outer-multiset} CA as those with
a local rule depending arbitrarily on a small central part of their
neighbourhood and on the multiset of other neighbouring states.
Formally, for any $k'$, $0 \leq k' \leq k$, $\oms{k'}{n,k}$ is the set of CA with $n$
states, arity $k$ and such that for any permutation $\pi$ of
${\{1\ldots k-k'\}}$ and any ${a_1,\ldots,a_{k-k'},
  b_1,\ldots,b_{k'}\in\Alphn}$ we have:
\begin{align*}
  \locA(a_1,\ldots,a_{\lfloor{(k-k')/2}\rfloor},& b_1,\ldots,b_{k'},a_{\lfloor{(k-k')/2}\rfloor+1},\ldots,a_{k-k'})\\
 & =\locA(a_{\pi(1)},\ldots,a_{\pi(\lfloor{(k-k')/2}\rfloor)},b_1,\ldots,b_{k'},a_{\pi(\lfloor{(k-k')/2}\rfloor+1)},\ldots,a_{\pi(k-k')}).
\end{align*}
We define in a similar way \emph{outer-totalistic} and
\emph{outer-set}, and denote
them by $\otot{k'}{n,k}$ and $\oset{k'}{n,k}$ respectively. Note that
what is classically called \emph{outer-totalistic} is exactly the
family $\otot{1}{n,k}$.

\focus{State symmetric CA} Families above are variations around the
invariance by permutations of neighbours. State symmetric CA are CA
with a local function invariant by permutation of the state
set. Formally, a CA ${\Ac\in\CAnk}$ is \emph{state symmetric}, denoted
by ${\Ac\in\SSnk}$, if for any permutation $\pi$ of $\Alphn$ we have:
\[\forall a_1,\ldots,a_k: \locA(a_1,\ldots,a_k) =
\pi^{-1}\bigl(\locA(\pi(a_1),\ldots,\pi(a_k))\bigr).\] Note that we
have a situation similar to the case of $\SET$: for fixed $k$, there
is a constant $K$ such that, for all $n$, ${\Size\SSnk\leq K}$. Thus their
is no hope that the asymptotic density of a non-trivial property for
fixed $k$ be $1$ in state symmetric CA.

\focus{Captive CA} Finally, we consider the family of captive CA
already introduced in \cite{captif}: they are CA where the local
function is constrained to produce only states already present in the
neighbourhood. Formally, a CA ${\Ac\in\CAnk}$ is \emph{captive},
denoted by ${\Ac\in\Knk}$, if:
\[\forall a_1,\ldots, a_k :
\locA(a_1,\ldots,a_k)\in\{a_1,\ldots,a_k\}.\] 
The following lemma
shows a strong relationship between captive and state symmetric CA.
\begin{lem}
  \label{lm:captot}
  Let $n$ and $k$ be such that ${1\leq k\leq n-2}$.  Then we have
  ${\SSnk\subseteq\Knk}$.
\end{lem}

\focus{Combining symmetries} In the sequel, we will often consider
intersections of two of the families above. Note that all
intersections are generally non-trivial. However, for the case of
$\TOTnk$ and $\Knk$, the intersection is empty as soon as there exists
two $k$-uple of states with disjoint support but with the same sum,
because the 'captive' constraint forces the two corresponding
transitions to be different whereas the 'totalistic' constraint forces
them to be equal. This happens for instance when ${n\geq 3}$ and ${k}$ is
even with $k$-uples ${(1,3,1,3,\ldots,1,3)}$ and ${(2,2,2,\ldots,2)}$.

\section{Simulations and Universality}
\label{sec:simuniv}

The property we are mostly interested in is intrinsic universality 
(see \cite{surveyOllinger} for a survey on universality). To formalize
it, we first define a notion of simulation.

A CA $\Ac$ is a \emph{sub-automaton} of a CA $\Ad$, denoted
${\Ac\sac\Ad}$, if there is an injective map $\varphi$ from $A$ to $B$
such that ${\overline{\varphi}\circ\globA=\globAd\circ
\overline{\varphi}}$, where ${\overline{\varphi}:A^\ZZ\rightarrow
B^\ZZ}$ denotes the uniform extension of $\varphi$ to configurations.
We sometimes write
${\Ac\sac_{\varphi}\Ad}$ to make $\varphi$ explicit. This definition is
standard but yields to a very limited notion of simulation: a given CA
can only admit a finite set of (non-isomorphic) CA as sub-automata.
Therefore, following works of J.~Mazoyer and I.~Rapaport \cite{gpcarr}
and later N.~Ollinger \cite{ollingerphd,surveyOllinger}, we will
consider the following notion of simulation: a CA $\Ac$ simulates an
AC $\Ad$ if some \emph{rescaling} of $\Ac$ is a sub-automaton of some
\emph{rescaling} of $\Ad$. The ingredients of the rescalings are
simple: packing cells into blocs, iterating the rule and composing
with a translation (formally, we use shift CA $\sigma_z$, $z\in\ZZ$, 
whose global
rule is given by $\sigma (c)_x = c_{x-z}$ for all $x\in\ZZ$). 
Formally, given any state set $Q$ and any $m\geq
1$, we define the bijective packing map ${\bloc{m}: Q^\ZZ\rightarrow
  \bigl(Q^m\bigr)^\ZZ}$ by:
\[\forall z\in\ZZ : \bigl(\bloc{m}(c)\bigr)(z) = \bigl(c(mz),\ldots,c(mz+m-1)\bigr)\]
for all ${c\in Q^\ZZ}$. The rescaling $\grp{\Ac}{m,t,z}$ of $\Ac$ by
parameters $m$ (packing), ${t\geq 1}$ (iterating) and ${z\in\ZZ}$
(shifting) is the CA of state set $Q^m$ and global rule:
\[\bloc{m} \circ \sigma_z\circ \globA^t \circ \debloc{m}.\]
With these definitions, we say that $\Ac$ simulates $\Ad$, denoted
${\Ad\simu\Ac}$, if there are rescaling parameters $m_1$, $m_2$,
$t_1$, $t_2$, $z_1$ and $z_2$ such that
${\grp{\Ad}{m_1,t_1,z_1}\sac\grp{\Ac}{m_2,t_2,z_2}}$.  In the sequel,
we will discuss \emph{supports} of simulations, \textit{i.e.} sets
of configurations on which simulations occur. If
${\grp{\Ad}{m_1,t_1,z_1}\sac_\varphi\grp{\Ac}{m_2,t_2,z_2}}$, the
support of the simulation is the set of configuration of $\Ac$ defined
by
${\debloc{m_2}\circ\overline{\varphi}\circ\bloc{m_1}(\AlphAd^\ZZ)}$. It
is a subshift: a closed shift-invariant set of configurations.
In the sequel we denote by $\Ad \simu_X \Ac$ the fact that $\Ac$ 
simulates $\Ad$ on support $X$.

Once formalised the notion of simulation, we naturally get a notion of
universality: CA able to simulate any other CA, denoted $\Ac\in \Univ$. This notion associated
to $\simu$ is called \emph{intrinsic universality} in the literature
(see \cite{surveyOllinger}). Actually, an intrinsically universal CA
$\Ac$ has the following stronger property (see
\cite{surveyOllinger,ollingerphd}): for all $\Ad$, there are rescaling
parameters $m$, $t$ and $z$ such that ${\Ad\sac\grp{\Ac}{m,t,z}}$.

\section{Asymptotic Density and Monotone Properties}
\label{sec:densprop}

\subsection{Asymptotic density} 

When considering a property $\Pro$ and a family $\Class$ (two sets of 
CA), we can define the probability of $\Pro$ in $\Classnk$ by $p_{n,k}=
\frac{\Size{\Classnk \cap \Pro}}{\Size{\Classnk}}$. Our probabilistic
framework consists in taking the limit of this probability $p_{n,k}$ 
when the "{size}" ($n$ and/or $k$) of the automata grows toward 
infinity. In \cite{Theyssier05}, only a particular case was 
considered: $k$ fixed, and $n\rightarrow \infty$. The following
definition consider all possible enumerations of '{size}'
through the notion of \textit{path}.

\begin{defi}
A path is an injective function $\Path: \NN \rightarrow \NN^2$. When 
the limit exists, we define the asymptotic density of $\Pro$ in 
$\Class$ following a path $\Path$ by $$\dlim{\Path}{\Class}{\Pro} 
  =\lim_{x\rightarrow \infty} \frac{\Size{\Class_{\Path(x)}\cap \Pro}}
  {\Size{\Class_{\Path(x)}}}$$
\end{defi}

The family of possible paths is huge and two different paths do not always define different densities. 

We denote $\N{c_0}= \Nn \setminus \{0,1,\ldots, c_0-1 \}$. Since we consider asymptotics, we can restrain to paths $\Path: \NN \rightarrow \N{n_0} \times \N{k_0}$ without loss of generality.

In the following, we will obtain limit densities of value $1$, which justifies the use of non-cumulative density : in our case a density $1$ following a given path implies a cumulative limit density $1$ along this path.

\subsection{Density of monotone properties among symmetric family}
A property $\Pro$ is said to be \emph{increasing} with respect to
simulation if $\forall \Ac \in \Pro$, $\Ac \simu \Ad$ implies $\Ad
\in \Pro$.  Decreasing properties are defined analogously. In this
section we prove that monotone properties have density $0$ or $1$
among symmetric families introduced in section \ref{sec:locsym}
following particular paths. More precisely, we are going to show that
any non-trivial increasing property has density 1.
\paragraph{}

For any local function $f:\Alphn^k \rightarrow \Alphn$, for any set 
$E \subseteq \Alphn^{k}$, we denote by $f|_{E}$ the restriction of $f$
to $E$. We also denote $\Classnk|_{E}=\{f|_{E}: f \in \Classnk\}$. Let
$\{E_i\}_{i\in I}$ be a finite family of subsets of $\Alphn^k$ and denote 
$E=\cup_{i\in I} E_i$. We say that the family $\{E_i\}_{i\in I}$ is 
independent for $\Class$ if the map 
$$\psi : \Classnk\ \rightarrow\  \Classnk|_{\Alphn^k \setminus E} \times \prod_{i\in I} \Classnk|_{E_i}$$
defined by
$ \psi(f) = (f|_{\Alphn^k \setminus E},f|_{E_1},\dots,f|_{E_i}, \dots)$
is a bijection (it is always injective).

By extension, we say that a collection of subshifts $\{X_i\}_{i\in I}$
is independent if the family $\{E(X_i)\}_{i\in I}$ is independent, 
where $E(X_i) \subseteq \Alphn^k$ is the set of words of length $k$ 
occurring in $X_i$. 

Let $\Simz=\{\Ac \in \CA : \Az \simu \Ac\}$ and $\Simzx{X}=\{\Ac \in \CA : \Az \simu_X \Ac\}$.

\begin{lem}
\label{lem:mapcoll}
Let $\Class \subseteq \CA$, and $\Az \in \Class_{n_0,k_0}$ a given CA.
For any size $(n,k)$ ($n\geq n_0$, $k\geq k_0$) and any collection of subshifts $\{X_i\}_{i\in I}$, we denote  $\alpha_i = \frac{\Size{\Classnk \cap \Simzx{X_i}}}{\Size{\Class_{n,k}}}$ for all $i$. If $\{X_i\}_{i\in I}$ is independent for $\Class$, then 
$$\frac{\Size{\Classnk \cap \Simz}}{\Size{\Class_{n,k}}}.
\geq 1- 
\prod_{i\in \Mapset} \left( 1- \alpha_{i}
\right)$$
\end{lem}

\begin{proof}
We use the notations above.
As the property $\Az \simu_{X_i} \Ac$ is only determined by the restriction of $\Ac$ to $E(X_i)$, there exists $A_i \subseteq \Classnk|_{E_i}$ such that $\psi(\Classnk\cap \Simzx{X_i}) = \Classnk|_{\Alphn^k \setminus E} \times \Classnk|_{E_1} \times \cdots \times \Classnk|_{E_{i-1}} \times A_i \times \Classnk|_{E_{i+1}}\cdots$.
And as the family $\{E_i\}_{i\in I}$ is independent for $\Class$, $\psi$ is bijective and $\alpha_i= \frac{\Size{A_i}}{\Size{\Classnk|_{E_i}}}$.

By definition of $\Simz$ we have the following inclusion: $\bigcup_{i\in I} (\Classnk \cap \Simzx{X_i}) \subseteq (\Classnk \cap \Simz)$. To conclude, it is enough to use the fact that $\psi$ is bijective in order to express the size of these sets' complement in $\Classnk$. 

\end{proof}

\vspace{-0.2cm}
\subsubsection{Increasing $n$, fixed $k$}

\begin{prop}
\label{lem:Galn}
In the following, $\ClassE$ is chosen among $\CA$, $\MS$, $\SET$, $\oset{k'}{}$, $\oms{k'}{}$.
For any $\Az \in \ClassE \cap \K_{n_0,k_0}$, for all $\epsilon$, there exists $n_{\epsilon,k_0}$ such that if $n\geq n_{\epsilon,k_0}$
$$\frac{\Size{(\ClassE \cap \K_{n,k_0}) \bigcap \Simz}}{\Size{\ClassE \cap \K_{n,k_0}}} \geq 1- \epsilon$$
\end{prop}

Thus, any increasing property $\Pro$ such that $\exists
\Az \in \ClassE \cap \K_{n,k} \cap \Pro$ has density $1$ in family 
$\ClassE \cap \K$ for paths with fixed $k$. The case $\ClassE=\CA$
was already proved in \cite{Theyssier05}.

\begin{proof}
Let $\{X_i\}_{i\in \ies{1}{\lfloor \frac{n}{n_0} \rfloor}}$ be a collection of 
fullshifts on disjoints alphabets of size $n_0$.
They are independent for family $\ClassE \cap \K$, whatever the choice of $\ClassE$.
Because of captivity constraint, the simulation happens on $X_i$
with probability $\alpha_{i,n,k_0} \geq c_0 ={1}/{n_0^{n_0^{k_0}}}$.
We obtain by lemma \ref{lem:mapcoll} $\frac{\Size{(\ClassE \cap \K_{n,k_0}) \bigcap \Simz}}{\Size{\ClassE \cap \K_{n,k_0}}} \geq 1- \left( 1- {c_0}\right) ^{\lfloor \frac{n}{n_0} \rfloor}$.

\end{proof}

\vspace{-0.2cm}
\subsubsection{Increasing $n$, fixed $k$}

In the following, we use lemma \ref{lem:mapcoll}, with an increasing
number $l=O(k)$ of independent simulation subshifts, each providing
the desired property for a constant fraction $d_n$ of
$\Class_{n,k}$ (n is fixed). It gives $ \frac{\Size{\Classnk \cap
    \Simz}}{\Size{\Classnk}} \geq 1- \left( 1- d_n \right)^{l}$ and we
obtain a limit density $\dlim{k}{\Class}{\Simz}=1$.

\focus{Multiset CA}

\begin{prop} For all $\Az \in \MS_{n_0,k_0}$, for all $\epsilon>0$,
for all $n\geq n_0+2$, there exists $k_\epsilon$ such that for all $k>k_\epsilon$, $
\frac{\Size{
\MS_{n,k} \cap \Simz}}{\Size{\MS_{n,k}}}>1-\epsilon$.
\end{prop}

\begin{proo} 
We consider a multiset CA $\Az \in \MS_{n_0,k_0}$, a size $n \geq n_0+2k_0+4$,
and a given $\epsilon >0$.
In order to clarify the construction we denote the $2$ biggest states of $\Alphn$ by $\0$ and $\1$.
For any size $k$, we define $l=\lfloor \frac{k-k_0}{k_0-1} \rfloor$ and $o=k-l.k_0$. And for any $j\in \ie{k_0+1}{l-k_0-1}$, $M_{j}$ is the word $M_{j}=\0^{l-j}\cdot\1^j$. 

We define the simulating subshift $X_j$ as the set of configurations alternating a state of $\Alphz$ and a pattern $M_{j}$. The family $\{X_j\}_j$ is independent for multiset CA.
On every such subshift, the simulation will happen if the CA maintains the structure (eventually shifted) and computes steps of $\Az$.
Multisets corresponding to patterns of length $k$ occurring in $X_j$ are:
\begin{itemize}
\item $V_{j,\ms{(x_1,1),(x_2,1),...,(x_{k_0},1)}}=$\\$ \ms{(\0,(k_0-1).j+o),(\1,(k_0-1).(l-j)),(x_1,1);(x_2,1),\ldots,(x_{k_0},1)} $
\item  For $0\leq s \leq o$, $W^0_{j,s,k_0-1}=$\\$ 
\ms{(\0,(k_0-1).j+o+1-s),(\1,(k_0-1).(l-j)+s),(x_1,1),(x_2,1),\ldots,(x_{k_0-1},1)} $
\item $W^1_{j,k_0-1}=
\ms{(\0,(k_0-1).j),(\1,(k_0-1).(l-j)+o+1),(x_1,1),(x_2,1),\ldots,(x_{k_0-1},1)}$
\item For $0\leq s \leq o-1$, $W^{1^{'}}_{j,s,k_0}=$\\$
\ms{(\0,(k_0-1).j+s),(\1,(k_0-1).(l-j)+o-s),(x_1,1),(x_2,1),\ldots,(x_{k_0},1)} $  

\end{itemize}

$\Az$ is simulated on support $X_j$ if we have the following:
\begin{itemize}
\item $\locA(V_{j,\ms{(x_1,1),(x_2,1),...,(x_{k_0},1)}}) = \locAz(\ms{(x_1,1),(x_2,1),...(x_{k_0},1)})$
\item $\locA(W^0_{j,s,k_0-1})=\0$ with $0\leq s \leq o$
\item $\locA(W^1_{j,k_0-1}) =\locA(W^{1^{'}}_{j,s,k_0})=\1$ with $0\leq s \leq o-1$
\end{itemize}

The number of involved legal multiset transitions for a given subshift $X_j$ is less than $(2.k_0+1).n_0^{k_0}$.
Thus, the proportion of CA in $\MSnk$  simulating $\Az$ on $X_j$ is at least $1/n^{(2.k_0+1).n_0^{k_0}}$ which is constant with increasing $k$.
And the number of such possible subshift is $l=O(k)$.
We conclude with lemma \ref{lem:mapcoll} as explained before.
\end{proo}

\focus{Totalistic CA}

We manage to make the multiset construction above
to become totalistic. To do it, we define the mapping $\varphi_j$ by: 
$\forall x \in \Alphz$, $\varphi_j(x)= (x(n_0+1))\cdot \0^{l-j}\cdot \1^j$, with $\0=0$ and $\1=n_0(n_0+1)+1$.
The $j$-th subshift is defined as the smallest subshift containing  $\left(\varphi_j(\Alphn^k)\right)^\ZZ$. The transitions are
distinguishable by the number of $\1$, and the number of states smaller than $n_0(n_0+1)$ in any legal neighbourhood. The
probability to simulate the original CA on the $j$-th subshift is
constant, and the simulating subshifts are independent for totalistic CA. As the number of possible
simulation increases, the limit probability for any CA to simulate a
given CA is increasing to $1$.

\focus{Outer-multiset CA}
We still consider the same \emph{possible} simulations of any multiset CA $\Az \in \CA_{n_0,k_0}$ by a CA $\oms{k'}{n,k}$.

As $\Ac$ is only partially multiset, the number of transitions involved in a simulation on one given subshift has increased: we have to consider the transitions with every possible central pattern of size $k'$.
Using a precise account, we ensure that the number of transitions involved in one given simulation is bounded by $c^{k'}$ with $c$ only depending on $n_0$ and $k_0$.
And the number of possible subshifts for the simulation to happen is the same as in the totally multiset case: it is
still given by $\lfloor k/2 \rfloor -1$. We obtain 
$\frac{\Size{\oms{k'}{n,k} \cap \Simz}}{\Size{\oms{k'}{n,k}}}>1- \left(1-\frac{1}{(n_0+2)^{c^{k'}}}\right) ^l$ with $l=O(k)$.
To ensure that $\dlim{k}{\oms{k'}{n,k}}{\Simz}=1$
it is enough to suppose that $k'= o(log(log(k))$.

\subsubsection{More general paths}

\focus{Multiset captive CA}
We prove a slightly more general result with the family of multiset
captive CA $\MSC$ defined by $\MSC= \K \cap \MS$.
\begin{prop} For any path $\Path: \NN \rightarrow \NN^2$ such that
  \emph{the lower limit of $x \mapsto n=\fst(\Path(x))$ is infinite},
  and for any $\Az \in \MSC_{n_0,k_0}$, for all $\epsilon$, there
  exists $s_{\epsilon}$ such that if $x>s_\epsilon$ then
$$\frac{\Size{\Simz \cap \MSC_{\Path(x)}}}{\Size{\MSC_{\Path(x)}}}>1-\epsilon$$
\end{prop}

\begin{proo} 
The collection of subshifts, and the simulation behaviour are exactly the same as in the multiset case. 
If $\Az$ is captive, each simulating transition is also captive.	
The number of involved transitions is the same as in the $\MS$ case: $(2.k_0+1).n_0^{k_0}$. But using the captivity constraint, the probability for the simulation on the $j$-th subshift to happen is also bounded by : $1/{(2.k_0+1).n_0^{k_0}}$.
We use the fact that the number of possible simulations is still $O(k)$ to conclude using lemma \ref{lem:mapcoll}.

\end{proo}

\focus{Set captive CA}

\begin{prop} For any path $\Path: \NN \rightarrow \NN^2$ such
that \emph{the lower limit of $x \rightarrow n=\fst(\Path(x))$  is infinite}, and for any $\Az \in \KSET_{n_0,k_0}$, for all $\epsilon$, there exists $s_{\epsilon}$ such that if $x>s_\epsilon$ then 
$$\frac{\Size{\Simz \cap \KSET_{Path(x)}}}{\Size{\KSET_{Path(x)}}}>1-\epsilon$$
\end{prop}

\begin{proo} Given $\Az$, $n$, and $k$ big enough, 
we denote the $2k_0+4$ first states of $\Alphn$ by $0_i$ and $1_i$, $i\in \ie{1}{k_0+2}$.
The $j$-th subshift is the set of configurations alternating words
$0_i^o1_i^{l-o}$ (with $l=\lfloor \frac{k-k_0}{k_0-1} \rfloor$ and $o=k-l.k_0$)
legally ordered and simulating states taken from $\Sigma_j=\ie{2k_0+4+j.n_0}{2k_0+4+j.n_0+n_0-1 }$. 
Legal set transitions for this subshift are
  \begin{itemize}
  \item $\{a_1,\ldots, a_{k_0}\} \cup \{\underline{0_i},1_i,\ldots,
0_{i+k-1},1_{i+k-1},\underline{0_{i+k}}\} \rightarrow
\locAz(\{a_1,\ldots, a_k\}) $
  \item $\{a_1,\ldots, a_{k_0+e}\} \cup \{\underline{1_{i-1}},0_i,1_i ,\ldots,
0_{i+k-1},1_{i+k-1},\underline{0_{i+k}}\} \rightarrow 1_{i+k/2} $ with $e\in \{0,-1\}$
  \item $\{a_1,\ldots, a_{k_0+e}\} \cup
\{\underline{0_{i}},1_{i},0_{i+1},1_{i+1},\ldots,
1_{i+k-1},0_{i+k},\underline{1_{i+k}}\} \rightarrow
0_{i+k/2} $ with $e\in \{0,-1\}$
  \end{itemize} 
With indicies modulo $k+2$, and $a_x \in \Sigma_j$ for all $x$.
For all $i$ those transitions may be identified by a set CA using the underlined state.
\paragraph{} So we need $n_0+2.(k_0+2)$ different states to make
the simulation on this subshift and the number of involved transitions is equal to $3.(k_0+2)$.  
Thus, because of captivity, the proportion $p$ of CA in which one given simulation
happens is constant when $k$, or $n$ is increasing.
And the family of the ${\lfloor \frac{n-2(k_0+2)}{n_0}\rfloor}$ possible simulation
subshifts is independent. With lemma \ref{lem:mapcoll}, we obtain the inequality
$\frac{\Size{\Simz \cap \KSETnk}}{\Size{\KSETnk}}>1-(1-p)^{\lfloor \frac{n-2(k_0+2)}{n_0}\rfloor}$.
  We conclude the proof using the hypothesis on the path, $\underline{lim}_{x\rightarrow \infty}n=\underline{lim}_{x\rightarrow \infty}\pi_1(x)=\infty$.
\end{proo}

\section{Encodings}
\label{sec:univexist}

In the following we prove that there exists universal cellular
automata in most of the families defined above. This is an important
step considering the fact that some well known locally defined family,
such as LR-permutative CA, do not contain any universal CA (because
intrinsic universality implies non-surjectivity,
see~\cite{ollingerphd}). In fact, for every given family $\Class$, we
introduce an encoding map ${\mapv_{\Class}:\CA\rightarrow\Class}$ such
that for any $\Ac$, its corresponding encoded version
$\mapv_{\Class}(\Ac)$ verifies
${\Ac\simu\mapv_{\Class}(\Ac)}$. The existence of a universal CA in
$\Class$ follows by application of the encoding to any universal CA.
Moreover, in some cases, we obtain a stronger result: the encoded CA
is universal if and only if the original CA is universal.

\focus{Set CA}

Given a CA $\Ac\in \CAnk$ of state set $\Alphn$, we construct
$\codms(\Ac)\in\SET$ with state set
${Q=\Alphn\times\{0,\ldots,k+1\}\cup\{\spr\}}$ of size ${n.(k+2)+1}$.

A configuration ${c\in Q^\ZZ}$ is said \emph{legal} if
${c(z)\not=\spr}$ for all $z$ and if the projection of $c$ on the
second component of states (which is well-defined) is periodic of
period ${1 \cdot 2\cdots (k+2)}$. Thus, for any legal configuration $c$
and any position $z$, the set of states of cells which are neighbours
of $z$ is of the form:
\[E_i(a_1,\ldots, a_k) = \{(a_1,i),(a_2,i+1\ mod\
k+2),...,(a_{k},i+k-1\ mod\ k+2)\}\] for some ${i\in\{1,\ldots,k+2\}}$
(with ${a_j\in\Alphn}$ for all $j$).  $\codms(\Ac)$ is defined by the
local rule $f$ as follows:
\[f(x_1,\ldots,x_k) =
\begin{cases}
  \bigl(\locA(a_1,\ldots,a_k),i+\lfloor k/2\rfloor\ mod\ k+2\bigr) &\text{if }\{x_1,\ldots,x_k\} = E_i(a_1,\ldots,a_k),\\
  \spr&\text{else.}
\end{cases}
\]

By construction, we have ${\codms(\Ac)\in\SET}$. Moreover the encoding preserves
universality. As a direct corollary, we get the undecidability of
universality in family $\SET$ (universality was proven undecidable in
the general case in \cite{Ollinger03}).

\begin{thm} 
  \label{thm:setuniv}
  The encoding ${\codms :\CA\rightarrow\SET}$ satisfies the following: 
  \begin{enumerate}
  \item ${\Ac\simu\codms(\Ac)}$ for all $\Ac$,
  \item $\Ac$ is universal if and only if $\codms(\Ac)$ is universal.
  \end{enumerate}
\end{thm}

\focus{Captive set CA} We denote by $\KSET$ the intersection
${\Kap\cap\SET}$. The previous construction does not generally produce
captive CA (even if the original CA is captive). We now describe a new
encoding which produces only CA belonging to $\KSET$. It could have
been used to prove the existence of universal set CA, but we have no
proof that it satisfies the second assertion of
theorem~\ref{thm:setuniv} (hence the usefulness of previous
construction).

The new mapping ${\mapv : \CA \rightarrow \KSET}$ is an adaptation
of $\codms$.  We keep the idea of states being a cartesian product of
the original alphabet $\Alphn$ and a family of labels which is in this
case $\{0,...,2k-2\}$.  But, in order to have every transition
satisfying the captive constraint, we introduce 'libraries' of states
placed regularly in legal configurations: between two computing
cells, we place the $i$-th library for some $i$, denoted by $\Li{i}$,
which contains the $n$ states $\{(x,i)\}_{x\in \Alphn}$.  For
technical reasons, it also contains special states $(\#,i)$ and
$(\#',i)$, and it is ordered as follows:
$\Li{i}=(\#,i)\cdot(1,i)\cdot(2,i)\cdots(n,i)\cdot(\#',i)$.  Thus, $\mapv(\Ac)$ has
state set ${Q = \{0,\ldots,2k-2\}\times(\Alphn\cup\{\#,\#'\})}$.

\newpage
The simulation of $\Ac$ by $\mapv(\Ac)$ takes place on 'legal'
configurations defined by an alternation of an isolated state of label
$i$, and a library of type $k+i$, precisely:
\[ \cdots\ (q_1,i)\ \cdot\ \Li{k+i \bmod 2k-1}\ \cdot\ (q_{2},{i+1 \bmod 2k-1})\
\cdot\ \Li{k+i+1 \bmod 2k-1}\cdots\] Those legal configurations will be
maintained in one-to-one correspondence with configurations of $\Ac$,
successive isolated states between libraries corresponding to
successive states from $\Ac$.  However, this time, the simulation of
$1$ step of $\Ac$ will use $2$ steps of $\mapv(\Ac)$ and only even
time steps of $\mapv(\Ac)$ (including time $0$) will produce legal
configurations. For odd time steps, we introduce 'intermediate'
configurations defined by an alternation of an isolated state of label
$i$, and a library of type $r+i$, precisely:
\[ \cdots\ (q_1,i)\ \cdot\ \Li{r+i \bmod 2k-1}\ \cdot\ (q_{2},{i+1
  \bmod 2k-1})\ \cdot\ \Li{r+i+1 \bmod 2k-1}\cdots\] where ${r=\lfloor
  k/2\rfloor}$ is the radius of $\Ac$.

To describe the local rule of $\mapv(\Ac)$, we introduce the
following sets:
\begin{itemize}
\item ${V_i(a_1,\ldots,a_k) = \{(a_1,i),(a_2,i+1\bmod 2k-1)\ldots
    (a_k,i+k-1\bmod 2k-1)\}}$;
\item $L_i$ is the set of states present in the word $\Li{i}$;
\item $B_{i,x} = \{(\#,i),(1,i),...,(b-1,i)\}$ is the set of states
  in the prefix of of $\Li{i}$ of length ${x}$;
\item $E_{i,x} =\{ (e,i),...,(n,i),(\#',i)\}$ is the set of states
  in the suffix of $\Li{i}$ of length ${n-x+1}$.
\end{itemize}

$\mapv(\Ac)$ has arity ${k'=k+(k-1)\cdot(n+2)}$ and, on legal
configurations, the set of states seen in a neighbourhood has one of
the following types:
\begin{description}
\item[T1] ${V_i(a_1,\ldots,a_k) \cup L_{i+k\ mod\ 2k-1} \cup\ldots \cup
    L_{i+2k-2\ mod\ 2k-1}}$;
\item[T2] ${V_i(a_1,\ldots,a_k)\cup E_{i+k-1\ mod\ 2k-1,x} \cup
    L_{i+k\ mod\ 2k-1} \cup \ldots}$\\ ${\ldots\cup L_{i+2k-3\ mod\ 2k-1} \cup
    B_{i+2k-2\ mod \ 2k-1,x}}$.
\end{description}

On intermediate configurations, the set of states seen in a
neighbourhood has one of the following types:
\begin{description}
\item[T3] ${V_i(a_1,\ldots,a_k) \cup L_{i-r\ mod\ 2k-1} \cup\ldots \cup
    L_{i-r+k-2\ mod\ 2k-1}}$;
\item[T4] ${V_i(a_1,\ldots,a_k)\cup E_{i-r-1\ mod\ 2k-2,x} \cup
    L_{i-r\ mod\ 2k-1} \cup \ldots}$\\ ${\ldots\cup L_{i-r+k-3\ mod\ 2k-1} \cup
    B_{i-r+k-2\ mod \ 2k-1,x}}$.
\end{description}

An important point is that the 4 types are disjoint: it is obvious
that each of T1 and T3 is disjoint from each of T2 and T4, and the
overall disjointness follows from the fact that sets of type T3 and T4
have less elements than T1 and T2 since set $L_i$ are disjoint
but \[V_i(a_1,\ldots,a_k)\cap L_{j}\not=\emptyset\iff i\leq j\leq
i+k-1\]

Using notations above, the behaviour of $\mapv(\Ac)$ is defined by 4
kinds of transitions according to the kind of neighbourhood seen:
\begin{description}
  \item[T1] ${\rightarrow (\locA(a_1,...a_{k}),i+2k-2\bmod 2k-1)}$
  \item[T2] ${\rightarrow (x,i+k-1 \bmod 2k-1)}$
  \item[T3] ${\rightarrow (a_{1+r},i+r \bmod 2k-1)}$
  \item[T4] ${\rightarrow (x,i+r)}$
\end{description}
The crucial point for transition of type T3 to be well-defined is that
$a_{1+r}$ can be unambiguously determined given that the libraries
present have labels ranging from ${i-r-1}$ to ${i-r+k-2 = i+r-1}$
whereas $a_{1+r}$ is associated to label ${i+r}$ in
${V_i(a_1,\ldots,a_k)}$ (everything is taken modulo ${2k-1}$).

Intuitively, type T1 corresponds to simulation of transitions of $\Ac$
and the three other types are devoted to the modification of label of
isolated states or the displacement of libraries according to the
following scheme:
\begin{itemize}
\item At even steps, transitions of type T1 apply the local rule
  $\locA$, but the result receive a label $j$ such that $\Li{j}$ is
  present in the neighbourhood to satisfy captivity constraint;
  meanwhile, transitions of type T2 just shift the libraries.
\item At odd steps, the difference of labels between libraries and
  isolated states is wrong; to come back to a legal configuration,
  transitions of type T3 leave isolated states unchanged while
  transitions of type T4 shift the libraries.
\end{itemize}

To completely define $\mapv(\Ac)$, we fix some ordering on $Q$ and
specify that, when the set $E$ of neighbours doesn't correspond to any
of the 4 types above, the local rule of $\mapv(\Ac)$ simply chooses the
greatest state in $E$.  With that definition, $\mapv(\Ac)$ always
belong to $\KSET$, because it depends only on the set of states in the
neighbourhood, and because each transition produces a state already
present in the neighbourhood (either the neighbourhood contains $L_i$
for the right value of $i$, or the local rule simply chooses the
greatest element).

\begin{thm}
  \label{thm:ksetuniv}
  For any $\Ac$, we have ${\Ac\simu\mapv(\Ac)}$. Therefore families
  $\MS$, $\MSK$, $\SET$ and $\KSET$ contain universal CA.
\end{thm}

The construction above corresponds to the strongest symmetry
constraint (captive set CA), put aside totalistic CA. The existence of
(intrinsically) universal totalistic CA is proven
in~\cite{ollingerphd}. The case of outer-totalistic CA follows by
inclusion.

\section{Universality Everywhere}
\label{sec:univdens}

Gathering the density results of section \ref{sec:densprop} and the existential results for universality in section \ref{sec:univexist}, we obtain an asymptotic density $1$ for universality in the following classes.

\begin{align*}
\begin{tabular}{|l|l|l|}
\hline
\textbf{Family $\Class$} &  \textbf{Condition on the path $\Path$} & \textbf{Comments}\\
\hline
Captive CA & $\exists k_0$ s.t. $\Path(x) =(x,k_0)$   & Already in \cite{Theyssier05}\\
Multiset CA &  $\exists n_0$ s.t. $\Path(x) =(n_0,x)$  &\\
$k'$-outer-multiset & $\exists n_0$ s.t. $\Path(x) =(n_0,x)$ &  $k'=o(\log(\log k))$\\
Totalistic CA & $\exists n_0$ s.t. $\Path(x) =(n_0,x)$  &\\
$k'$-outer-totalistic & $\exists n_0$ s.t. $\Path(x) =(n_0,x)$ & $k'=o(\log(\log k))$ \\
Set captive CA &  $\underline{\lim}_{x\rightarrow \infty}\ \fst(\Path(x))= +\infty $ & \\
Multiset captive CA &  $\underline{\lim}_{x\rightarrow \infty}\ \fst(\Path(x))= +\infty $ &\\
\hline
\end{tabular}
\end{align*}

\vspace{0.2cm}

\section{Open Problems and Future Work}
\label{sec:open}

As summarised in the previous section, our work establishes that
universality has asymptotic density 1 along path $\Path$ in several
families defined by local symmetries, provided $\Path$ verifies some
hypothesis depending on the family considered.

Notably, we leave open the question of the density of universality in
the following cases:
\begin{itemize}
\item increasing state set for families $\MS$, $\SET$, $\TOT$ (and
  outer-versions),
\item increasing neighbourhood for family $\Kap$.
\end{itemize}

We have no result (and no intuition) concerning the case of the whole
set of CA either. A possible progress on that topic could be to reduce
the density problem of a family $\Class_1$ to the density problem of a
family $\Class_2$, \textit{i.e.} to show that the densities (if they
exist) in the two families are equal up to non-trivial multiplicative
constants.

Another perspective, especially for multiset CA (or sub-families
$\SET$ and $\TOT$), is to extend our result to higher dimensions or
even to more general lattice of cells. Indeed, the symmetry involved
here implies isotropy which is an often required property in
modelling.

Finally, it remains to study typical dynamics obtained in each family
from random initial configuration. Experiments suggest that
self-organisation in those families is far more frequent than in CA in
general.

\bibliographystyle{plain}
\bibliography{ac}

\end{document}